\documentclass[pre, aps, twocolumn, floatfix, showkeys, superscriptaddress, nofootinbib]{revtex4-1}
\usepackage{amsmath}
\usepackage{booktabs}
\usepackage{bm}
\usepackage{hhline}
\usepackage{graphicx}
\usepackage{xcolor}

\newcommand{\br}{{\bm r}}
\newcommand{\brp}{{\bm r}^\prime}

\newcommand{\bx}{{\bm x}}
\newcommand{\bxp}{{\bm x}^\prime}

\newcommand{\bcr}{{\bm R}}

\begin{document}

\title{Effect of Ionic Disorder on the Principal Shock Hugoniot}
\author{Crystal F. Ottoway}
\author{Daniel A. Rehn}
\author{Didier Saumon}
\author{C. E. Starrett}
\email{starrett@lanl.gov}

\affiliation{Los Alamos National Laboratory, P.O. Box 1663, Los Alamos, NM 87545, U.S.A.}
\date{\today}
\begin{abstract}
   The effect of ionic disorder on the principal Hugoniot is investigated using Multiple Scattering Theory to very high pressure (Gbar).  Calculations using molecular dynamics to simulate ionic disorder, are compared to those with a fixed crystal lattice, for both carbon and aluminum.  For the range of conditions considered here, we find that ionic disorder is most important at the onset of shell ionization and that at higher pressures, the subtle effect of the ionic environment is overwhelmed by the larger number of ionized electrons with higher thermal energies.  
\end{abstract}
\pacs{}
\keywords{warm dense matter, Hugoniot, equation of state}

\maketitle

\section{Introduction}
The principal shock Hugoniot is the locus of final states in single-shock experiments where the material is initially at standard temperature and pressure.   It is a method with a long history \cite{marsh1980lasl}, and it is used to measure the equation of state at high compressions and temperatures.  In flyer plate shock experiments, a pusher is driven into the material of interest, launching a shock wave.  In laser-driven shock experiments, a layer of ablator material is laser-heated.  The rapid expansion of the ablator launches a shock wave into the target material.  Conservation of mass, energy and momentum across the shock front, and the assumption of an ideal shock, translate the measured pusher and shock front velocities into the desired equation of state variables.

Recent experiments at the National Ignition Facility (NIF) \cite{lindl04nif}, have resulted in accurate measurements of the Hugoniot curve at very high pressures \cite{kritcher2020measurement}, and are of relevance to inertial confinement fusion (ICF) and white dwarf physics \cite{dufour2007white}.  These experiments at 100's of Mbar complement the older, lower pressure techniques, including gas-gun \cite{nellis1983equation}, Diamond Anvil Cells \cite{mao94ultra}, and lower energy laser compression \cite{cauble98absolute, doppner18absolute}, that have challenged and guided EOS models and tables for many years \cite{lyon1992sesame}.

Models for the equation of state have for many years used these experiments for validation and testing.  Practical models, used to build equation of state tables, need to be reasonably accurate, computationally cheap, and applicable over a huge range of conditions and materials.  These pragmatic restrictions often require the use of simplified models of the EOS physics.  For example, many widely used EOS tables are based on so-called average atom models.  These attempt to capture the properties of one averaged atom that is representative of the system \cite{feynman49,rozsnyai72relativistic, piron11, liberman, wilson06, starrett2019wide}, but do not include the effect of ionic disorder.  As a result, ionic disorder has to be included via a separate model \cite{more88quot,johnson91generic, burnett18sesame}.

The effect that a consistent treatment of ionic disorder has on Hugoniot curves remains largely untested for dense plasmas.  Some modeling methods are capable of evaluating this.  For example, the widely used Density Functional Theory Molecular Dynamics (DFT-MD) method includes this ionic disorder through the use of ensemble averaging over MD time steps \cite{collins95, desjarlais03density}.  While accurate, this method is generally limited to degenerate systems (read lower temperature, higher density) due to computational expense.  Another method that includes ionic disorder is Path Integral Monte Carlo (PIMC) \cite{militzer_fpeos}.  This is also an accurate method, but is generally restricted to high temperatures due to the Fermion sign problem \cite{militzer_fpeos}.

Recently, a DFT based method that can reach high temperatures has been developed.  This method, known as Multiple Scattering Theory (MST), has a long history in solid state physics \cite{korringa47, kohn54}.  It has recently been adapted it to high temperature, dense plasmas \cite{starrett_2020_ms_thoery_for_plasma, laraia2021real}.  MST includes a sophisticated DFT treatment of the electrons, and includes ionic disorder through ensemble averaging over ionic configurations obtained with molecular dynamics simulations.  

In this work, we use MST to assess the impact of ionic disorder on the principal Hugoniots of carbon and aluminum.  We report results up to several Gbar using both ionic configurations from molecular dynamics and for a fixed crystal lattice structure.  We also compare to an average atom model and existing ab initio simulations \cite{militzer_fpeos}.  We find that inclusion of ionic disorder has a larger effect for aluminum than carbon, and that at high pressures, where the plasma is significantly ionized, a crude treatment of ionic disorder is accurate enough for Hugoniot predictions.

\section{Method}

Multiple Scattering Theory (MST) is based on the idea that the time-independent Green's function for a system containing electrons and nuclei can be found by solving Dyson's equation.  This equation relates the a priori unknown Green's function $G(\bx,\bxp, \epsilon)$ to the known Green's function of some reference system $g(\bx, \bxp, \epsilon)$ \cite{herrera2021greens}
\begin{multline}
     G(\bx,\bxp,\epsilon) = g(\bx,\bxp,\epsilon) \\+ \int d{\bx_1}G(\bx,\bx_1,\epsilon) V(\bx_1) g(\bx_1,\bxp,\epsilon)
\end{multline}
where $V(\bx)$ is the potential difference between the reference system and the desired electron-nucleus system.  Choosing a free-electron reference system,
\begin{equation}
    g(\bx,\bxp,\epsilon) = -\frac{m_e}{2\pi} \frac{\exp\left( \imath \mid \bx - \bxp \mid \right) } {\mid \bx - \bxp \mid}
\end{equation}
where $m_e$ is the electron mass, $V(\bx)$ becomes the potential for the electron-nucleus system.

The next step is to carry-out a multi-center expansion of the Green's functions using spherical harmonics $Y_{lm}(\hat{\br})$.  The positions of the expansion centers were originally chosen to coincide with the nuclear positions \cite{korringa47,kohn54}.  This works well for close-packed crystal structures.  However, it is not appropriate for disorder plasmas, and in addition to these nuclear centers, extra expansion centers are added \cite{starrett_2020_ms_thoery_for_plasma}.  These expansion centers are used to tessellate space into space-filling, non-overlapping cells.  The result of this multi-center expansion is
\begin{multline}
  G(\br+\bcr^n ,\brp+\bcr^{n'},z)= G^{ss}(\br+\bcr^n ,\brp+\bcr^{n'},z) \\
  + G^{ms}(\br+\bcr^n ,\brp+\bcr^{n'},z)
  \label{gfa}
\end{multline}
where $\bcr^n$ is the position vector of the $n^{th}$ expansion center, $\br$ is a vector pointing from this center to a point within cell $n$, and $z$ is a (in general, complex) electron energy.  The so-called single-site Green's function $G^{ss}$ is
\begin{multline}
  G^{ss}(\br+\bcr^n ,\brp+\bcr^{n'},z)= \\
  2m_e\delta_{nn'}
  \sum_{L=0}^\infty H^{n,\times}_L(\br_>,z) R^{n}_L(\br_<,z)
  \label{gfss}
\end{multline}
where $L=\{l,m\}$, i.e., the usual orbital angular momentum and magnetic quantum numbers, $H_L^n(\br,z)$ and $R_L^n(\br,z)$ are the irregular and the regular solutions of the Schr\"odinger equation, and the notation $\br_>$ ($\br_<$) means to take $\br$ or $\br'$ according to which one is greater (lesser) in magnitude.  This single-site Green's function corresponds to the Green's function for a cell with free-electron boundary conditions.  The so-called multi-site Green's function is
\begin{multline}
  G^{ms}(\br+\bcr^n ,\brp+\bcr^{n'},z)= \\
  2m_e\sum_{LL'}^\infty R^{n}_L(\br,z)
  {\cal G}^{nn'}_{LL'}(z)   
  R^{n'\times}_{L'}(\brp,z)
  \label{gfms}
\end{multline}
which can be viewed as a correction to the single-site Green's function that modifies the boundary conditions such that all incoming and outgoing waves from the cells match at the interfaces.
The superscript $\times$ means to take the complex conjugate of the angular part of $H^n_L$ or $R^n_L$.
Throughout, we use Hartree atomic units with $\hbar = 4\pi\epsilon_0 = e^2 = 1$, leaving $m_e$ symbolic for easy conversion to Rydberg units.
For other normalization and sign conventions, see reference~\cite{starrett_2020_ms_thoery_for_plasma}.

The ${\cal G}^{nn'}_{LL'}(z)$ are elements of the so-called structural Green's function matrix $\bm{\mathcal{G}}(z)$.  This is found by solving a variation of Dyson's equation, in what is sometimes known as the fundamental equation of MST,
\begin{equation}
  {\bm{\mathcal G}}(z)=
  {\bm{\mathcal G}}_0(z)
  \left[I-
    {\bm{t}}(z)
    {\bm{\mathcal G}}_0(z)
  \right]^{-1}
  \label{sgf}
\end{equation}
Here $\bm{t}(z)$ is the t-matrix, found by matching the numerical solutions $R^n_L$ and $H_L^n$ to their free electron forms at the cell boundaries \cite{zabloudil_book}.  ${\bm{\mathcal G}}_0(z)$ is the structure constants matrix.  Its dependence on the set $\{\bcr_n\}$ has been suppressed in the notation.  For a given set of expansion centers and energies, it can be calculated using a cluster approximation \cite{laraia2021real} or assuming a periodically repeating crystal structure \cite{starrett_2020_ms_thoery_for_plasma}.  Here we use the cluster approximation for all plasma conditions.  For the initial state calculations for the Hugoniots (see section \ref{sec3}) we use the periodic crystal structure calculation.  

\begin{figure}[h]
\centering
\includegraphics[scale=.54]{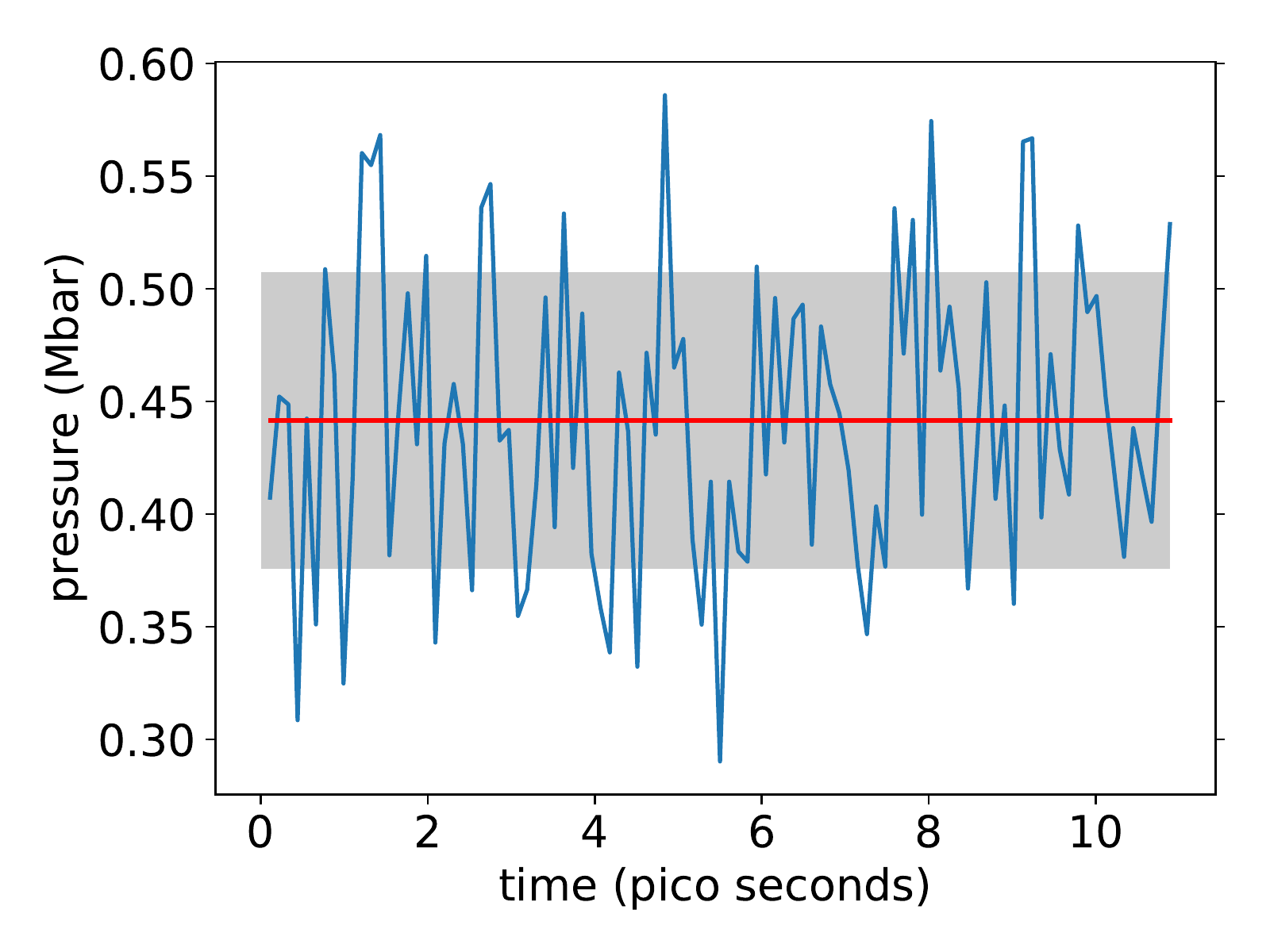}
\caption{Time variation of pressure for aluminum at 1 eV and 2.7 g/cm$^3$.  The pressure fluctuates over molecular dynamics time steps (blue line).  The average pressure is shown by the red horizontal line, and the grey area covers one standard deviation from this mean value.  For this case, the time step is 2.2 fs, and we calculate the pressure at every fiftieth time step.}
\label{fig:ave_over_frames}
\end{figure}

In practice these equations are solved using Mermin-Kohn-Sham density functional theory \cite{kohn65, mermin65}.  For a given set of expansion centers, the t-matrix, regular and irregular solutions to the Kohn-Sham equation are found for all cells, and the Green's function is then constructed.  Note that the global t-matrix ${\bm t}(z)$ is block diagonal.  Each block element corresponds to a t-matrix for a particular cell.   In the calculations and results presented here we use the muffin-tin approximation, where the effective Kohn-Sham potential in each cell is spherically averaged inside the muffin-tin radius, and takes a constant interstitial value elsewhere, as detailed in \cite{starrett_2020_ms_thoery_for_plasma}.  Further, we have used the temperature dependent Local Density Approximation (LDA) of Karasiev et al \cite{karasiev14} for the exchange and correlation functional.  The nuclear positions are provided by an external model: PseudoAtom Molecular Dynamics (PAMD) \cite{starrett15a}, which is thought to be accurate for all materials and conditions considered here.  The equation of state is then calculated as a time average over uncorrelated time steps (figure \ref{fig:ave_over_frames}).

\begin{figure*}[h]
\centering
\includegraphics[scale=.45]{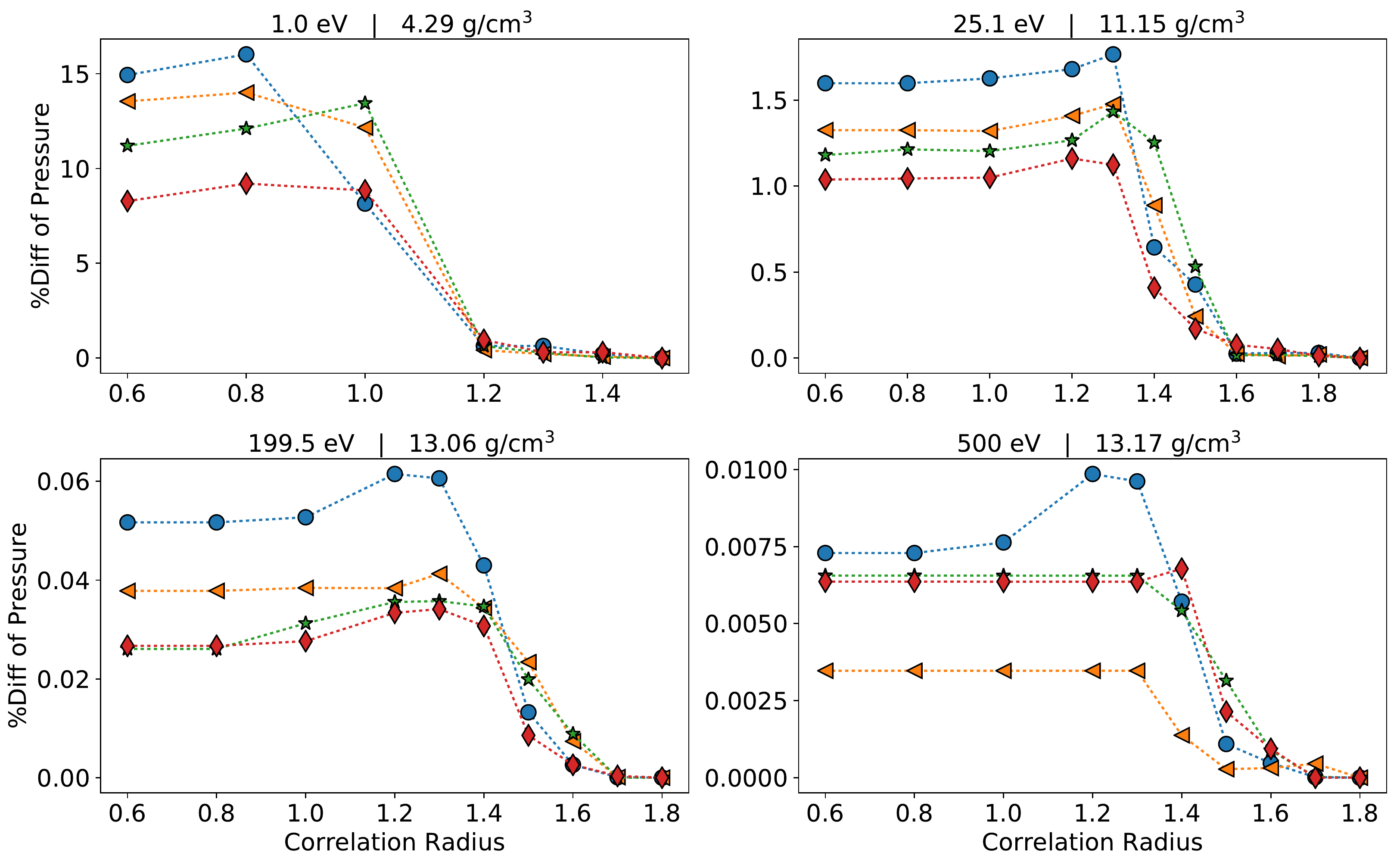}
\caption{The effect of the correlation radius on the pressure, for aluminum plasmas at conditions close to points on the principal Hugoniot.  Each panel corresponds to a different temperature and density.  Each line corresponds to a different molecular dynamics snapshot.  The percent change in the pressure is relative to the value at the largest correlation radius shown.}
\label{fig:al_rc_conv_pres_full}
\end{figure*}
\begin{figure*}[h]
\centering
\includegraphics[scale=.45]{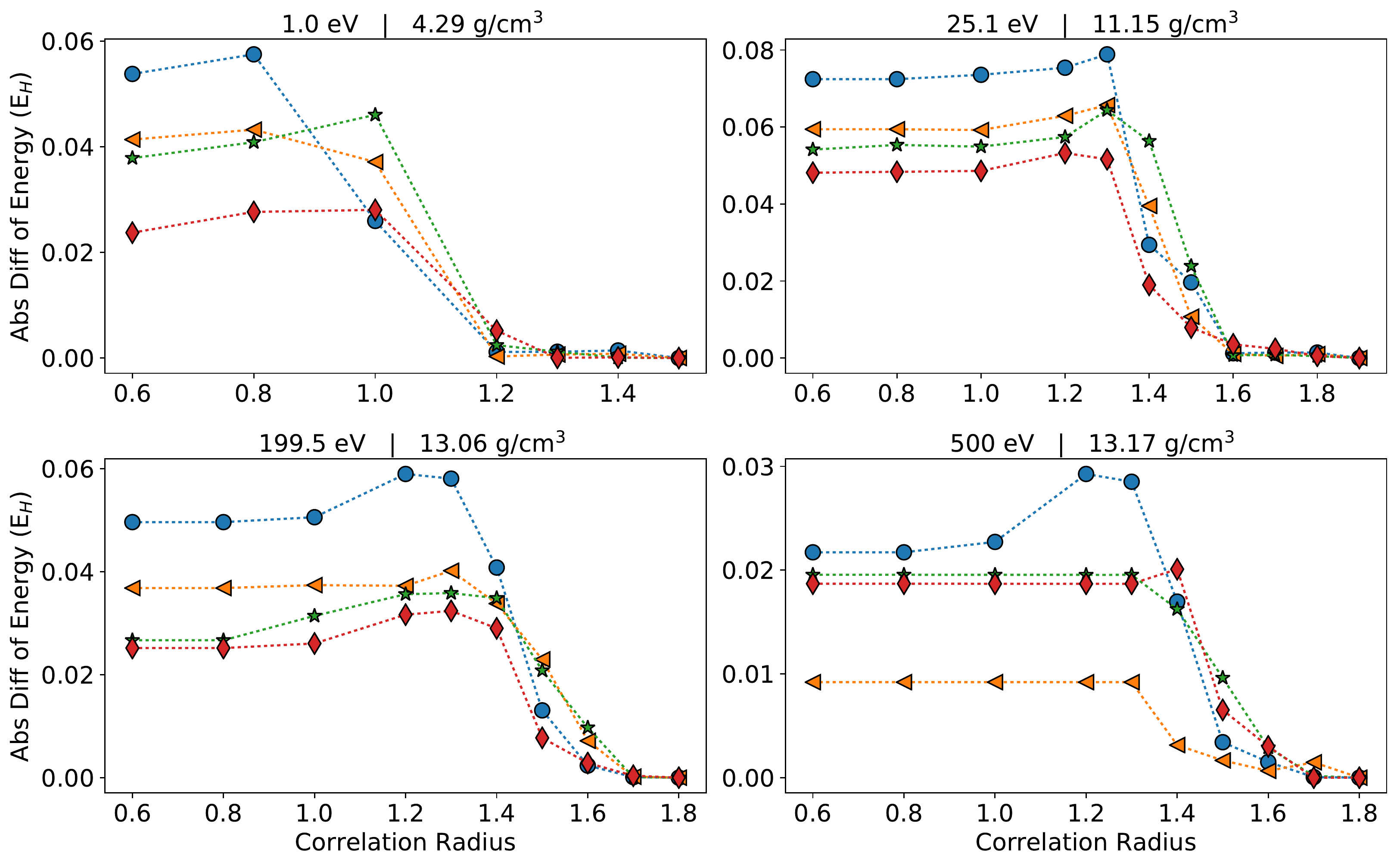}
\caption{The effect of the correlation radius on the internal energy per atom, for aluminum plasmas at conditions close to points on the principal Hugoniot.  Each panel corresponds to a different temperature and density.  Each line corresponds to a different molecular dynamics snapshot.  The absolute change in the energy is relative to the value at the largest correlation radius shown.}
\label{fig:al_rc_conv_ener_full}
\end{figure*}
The infinite sum over $L$ in equation \ref{gfss} is in practice converged automatically, and only a finite number of terms are needed \footnote{The number depends on the degeneracy of the system.  For cold, dense material, only a few terms are needed, while for hot dilute systems, many are needed}.  The two infinite sums in equation \ref{gfms} are more challenging due to high computational expense.  In practice, we use the method analysed in reference \cite{starrett_2020_ms_thoery_for_plasma}, where only `chemically relevant' terms are retained.  Here we keep terms up to and including $l=2$, which should be sufficient for the cases considered here.  Calculations with higher numbers of terms were considered in reference \cite{starrett_2020_ms_thoery_for_plasma} without significant changes to the EOS.
 
With the Green's function determined, the electron density is calculated
\begin{equation}
n_e(\br) = -\frac{2}{\pi}\Im \int_{-\infty}^{\infty}d\epsilon f(\epsilon,\mu)G(\br,\br,\epsilon)
\label{ne_gf}
\end{equation}
where the $2$ is due to spin degeneracy, the integral is along the real energy axis, and $f$ is the Fermi-Dirac function with chemical potential $\mu$.  A key advantage of the MST method is that, because the (retarded) Green's function is analytic in the upper-half complex energy plane, the energy integral in equation (\ref{ne_gf}) can be carried out using Cauchy's integral theorem \cite{ebert11,mavropoulos2006korringa}.  On the real energy axis, the full structure of the Green's function must be resolved.  For periodic systems, this means that one would need to find all the discrete eigenvalues, a notoriously difficult problem \cite{ebert11}.  By carrying out the integral in the complex energy plane, this problem is completely avoided - one no longer solves an eigenvalue problem.  Moreover, the integrand becomes a smooth function of the energy away from the real energy axis \cite{johnson84}, reducing the number of quadrature points needed.

The equations are then solved to self-consistency.  We use Eyert's acceleration method, which is a quasi-Newton technique, to speed up convergence \cite{eyert96}.  The equation of state can then be calculated using the method given in reference \cite{starrett_2020_ms_thoery_for_plasma}.  The initial guess for the potential in each cell is based on the Thomas-Fermi cell model \cite{feynman49}.

\section{Results\label{sec3}}
 As in reference \cite{laraia2021real}, our physical model is a computational cube that is periodically repeated.  As discussed above, a cluster approximation is used to solve equation \ref{sgf}.  The cluster should contain enough centers in it such that adding more does not affect the desired quantities.  Our cluster contains, at a minimum, the centers in the computational cube.  We then add centers within a fixed distance, called the correlation radius, of any center in the computational cube.  Hence, for zero correlation radius the cluster includes all centers within the computational volume, which for the cases presented here is 43 centers (8 nuclei + 35 extra centers). 
In figures \ref{fig:al_rc_conv_pres_full} and \ref{fig:al_rc_conv_ener_full} we show the the effect of increasing cluster size on the pressure and energy for density and pressure points close to the Hugoniot curve for aluminum. Figure \ref{fig:al_rc_conv_pres_full} shows that for pressure, relative errors of less than 1\% are achieved at 1 eV and 4.29 g/cm$^3$ with a correlation radius of 1.2 ion-sphere radii, and at the highest temperature case, much smaller relative error is seen for all correlation radii.  In between these extremes, the relative error steadily decreases as temperature increases.  This behavior can be explained by noting that the relative contribution of the multiple scattering Green's function becomes smaller as temperature increases.  This is because the scattering electrons have more energy on average, and are therefore more free-electron like.  As such, the correction to the single site term of the Green's function becomes smaller as temperature increases.  This effect was already noted in references \cite{starrett_2020_ms_thoery_for_plasma} and \cite{laraia2021real}.

For the internal energy, figure \ref{fig:al_rc_conv_ener_full}, somewhat different trends are observed.  We show the absolute difference in energy from the largest correlation radius for each case.  Similarly sized errors are observed for all temperatures, except the highest temperature where errors are roughly a factor of two smaller.  To explain this, we first note that we show absolute sizes of errors, which are more meaningful for energies as energy is not on a unique scale.  The figure shows that the effect of multiple scattering on the energy is fairly constant.  This implies that the number of electrons in states that are sensitive to multiple scattering (i.e. states with energies close to the threshold required for an electron to escape a cell), remains fairly constant over the temperatures and densities considered, and hence contribute similarly to the total energy.  The number of electrons in such states will result from a complicated interplay of effects.  A decreased electron degeneracy (as temperature increases) could either increase or decrease the number of such electrons though a reduction in Fermi-Dirac occupation factors of near threshold states, and an increase in ionization. Pressure ionization, where core states become valence states, could lead to an increase in the number of these electrons.

 \begin{table*}
\begin{center}
    \bgroup
    \begin{tabular}{c c c c c c}
 \hline
  Temperature (eV) &Density (g/cm$^3$) & \multicolumn{2}{c}{Pressure (Mbar) }  &abs(diff) &\%diff \\
                   &                  & 8+30 centers & 8+35 centers &     & \\
 \hline
10 & 3.52 & 10.205 $\pm$ 0.604 & 10.228 $\pm$ 0.627 & 0.0232 & 0.23\%
\\
100 &3.52 &140.038 $\pm$ 0.953 &139.916	$\pm$ 1.084 &0.1222 &0.09\%
\\ 
\end{tabular}
\egroup
\end{center}
\caption{The effect of the number of extra centers on the pressure of carbon with 8 nuclei.  No significant difference is observed at these conditions between 30 and 35 extra centers.  We used 35 extra centers for all MD calculations presented, having checked convergence at other conditions.\label{tab_ex_cent}}
 \end{table*}

In table \ref{tab_ex_cent} we show the effect of the number of extra centers on the pressure for carbon plasmas at two temperatures.  We have previously explored this effect in reference \cite{laraia2021real} for aluminum for particular molecular dynamics frames.  Here we show the time averaged pressure and standard deviation for a computational box with 8 unique nuclei, for both 30 and 35 extra centers.  For both temperatures the pressure is not significantly affected by the difference in the number of extra centers.  For reference, we have used 8 nuclei and 35 extra centers, and 100 molecular dynamics frames, for all plasma calculations presented here.  

It is worth commenting on whether this number of particles is sufficient for present purposes.  Unfortunately we cannot definitively answer this question quantitatively, as the memory requirement of our code currently limits calculations to these small systems.  Based on other studies \cite{danel2015equation}, and our own experience, we expect that larger system sizes will be necessary for more strongly correlated ionic fluids, i.e., lower temperatures and higher densities.  For the principal Hugoniot, therefore, the points most effected will be those at low-temperature, near the initial conditions.  To mitigate this issue, we therefore only calculate Hugoniot points above 1 eV ($\approx 11600$ K).

The Hugoniot curve is the solution to the Rankine-Hugoniot equation
\begin{equation} \label{eq:8}
    \frac{1}{2}(v_0-v)(P+P_0)- (e - e_0) = 0
\end{equation} 
where $v$ is the specific volume, $e$ the specific internal energy, and $P$ the pressure.  The subscript $0$ refers the those quantities of the initial, unshocked, state of the material.  Given an equation of state $P=P(v,T)$, $e=e(v,T)$, equation \ref{eq:8} has a sequence of solutions (the Hugoniot), that can be expressed as $T(v)$.  For an initial state, the resulting locus of final states in a single shock experiment is the principal Hugoniot.

\begin{figure}
\centering
\includegraphics[scale=.37]{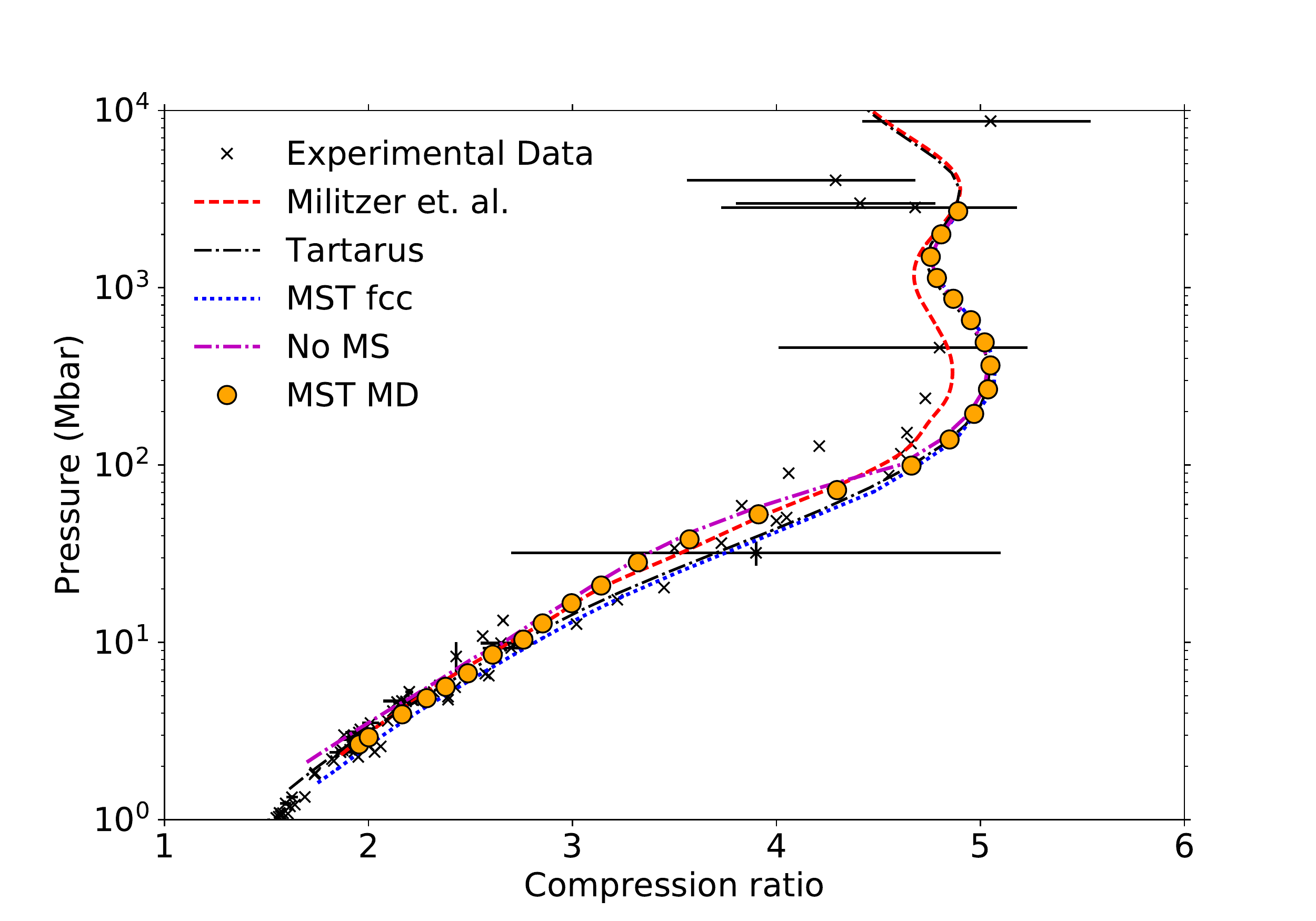}
\caption{Aluminum Hugoniot with experimental data \cite{altshuler_1974,altshuler_1961,altshuler_1977,avronin_1986, simonenko_1985, glushak_1989, isbell_1968, kormer_1962, mcqueen_1970, mitchell_1991, trunin_1995, ragan_1982, ragan_1984, vladimirov_1984, volkov_1980,model_1985}.  
Shown are results from the present model with molecular dynamics configurations (MST MD), the present model with an fcc lattice (MST fcc), the \texttt{Tartarus} average atom model \cite{starrett2019wide}, as well as results from the FPEOS of Militzer et al \cite{militzer_al, militzer_fpeos}.  The ``No MS" calculation is a simplified model described in the text.  Note that the initial density was taken to be 2.7 g/cm$^3$.}
\label{fig:al_hug_w_exp_data}
\end{figure}

In figure \ref{fig:al_hug_w_exp_data} the principal Hugoniot for aluminum is shown.  
Let us first consider the comparison of our full model calculation labelled MST MD which includes disordered, molecular dynamics determined, nuclear positions.  The curve labeled MST fcc uses the same MST method but assumes an fcc crystal structure at all temperatures and densities.  The initial state was calculated assuming a fcc structure for the structure constants, and it is same for both calculations.  The differences between the calculations are then solely due to the treatment of the ionic disorder.  

We see that overall, these calculations are in reasonable agreement, especially at high pressures, above 100 Mbar, corresponding to temperatures greater than 40 eV.  This high pressure region is where the ionization of the $n=2$ (lower pressure lobe) and $n=1$ (higher pressure lobe) shells occur.  That ionic structure does not strongly influence these features is due the increased thermal energy of the ionized electrons.  They are more free-electron like and are therefore not sensitive to the ionic structure.

At lower pressures (12 - 100 Mbar) the calculation with ionic disorder (MST MD) predicts that the plasma is stiffer (less compressible) than the fcc calculation (MST fcc).  In this pressure region, we find that both pressure and energy increase relative to the fcc calculation.  At a given temperature, an increase in energy leads to a more compressible Hugoniot, while an increase in pressure leads to a stiffer, less compressible Hugoniot.  Thus, the overall effect on the Hugoniot depends on a delicate interplay of these competing effects. 

\begin{figure}
\centering
\includegraphics[scale=.37]{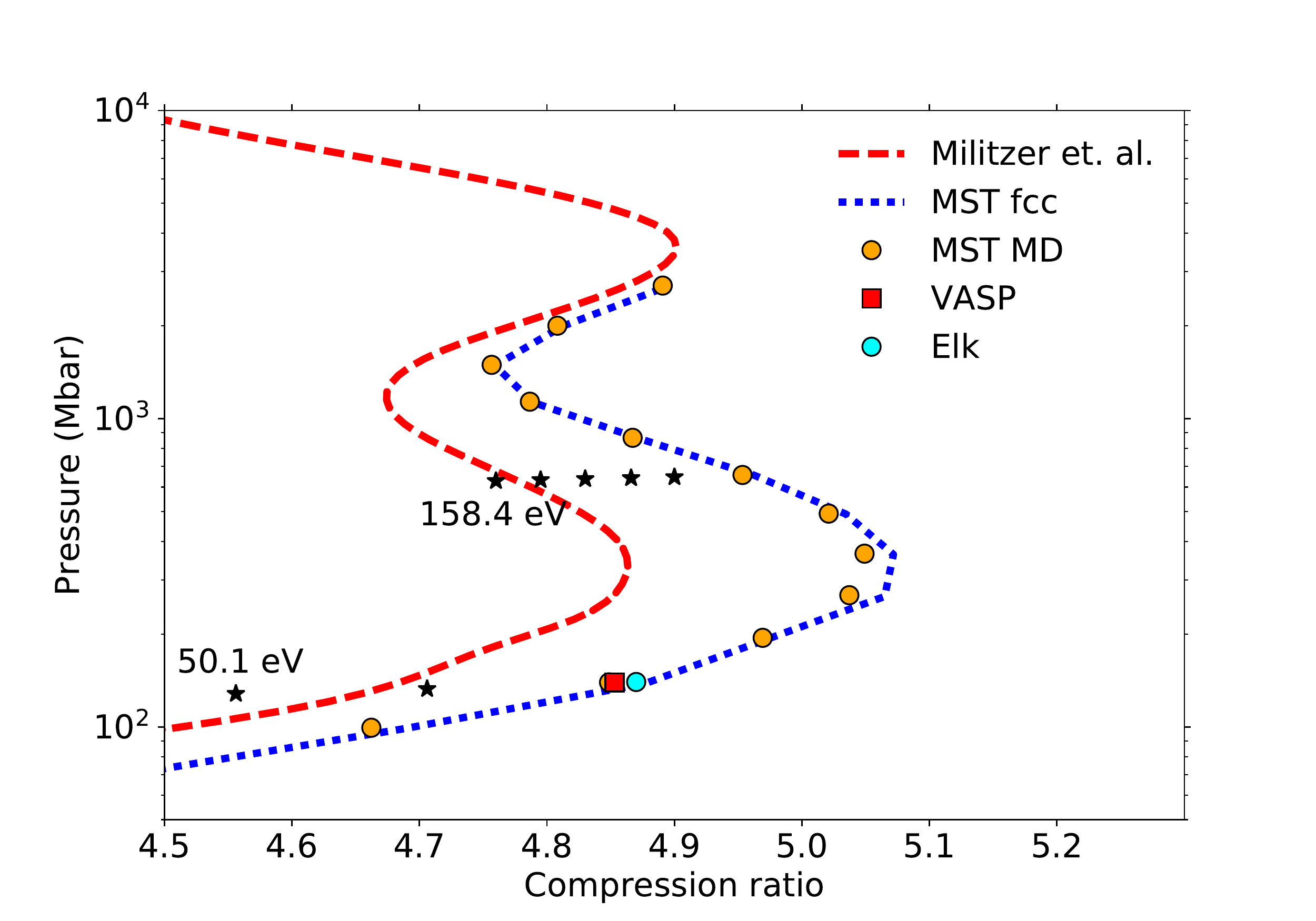}
\caption{Aluminum Hugoniot focused on the region of ionization of the $n=2$ shell.  Shown are the present multiple scattering calculation with an fcc lattice (MST fcc), and the present model with molecular dynamics configurations (MST MD), as well as results from the FPEOS of Militzer et al \cite{militzer_al, militzer_fpeos}.  The points labeled ``Elk'' and ``VASP'' are our own calculations of a Hugoniot point at 50.1 eV using the Elk \cite{elk} and VASP  \cite{kresse1993ab,kresse1996efficient,kresse1996efficiency} codes, where an fcc lattice has been assumed.
The black stars show the effect of reducing the MST internal energy in steps of 1 E$_H$ (right to left stars, starting at 1 E$_H$).  The two set of crosses correspond to the temperatures labeled on the plot.
}
\label{fig:al_hack}
\end{figure}

Also show in figure \ref{fig:al_hug_w_exp_data}, is the result from the average atom model {\texttt Tartarus} \cite{starrett_tartarus}, which does not include ionic disorder.  From that point of view, it is even simpler than the MST fcc model, but is similar in spirit.  The agreement between this model and the fcc calculation seems to bear this out.  Some differences between these two models appear at low compressions and pressures.  Clearly, the MST fcc calculation will give a much more realistic prediction of the initial state \cite{starrett18} than the average atom model, leading to the differences observed in the figure.  Unfortunately the experimental data, also shown in the figure, does not discriminate between the models.

\begin{figure}
\centering
\includegraphics[scale=.55]{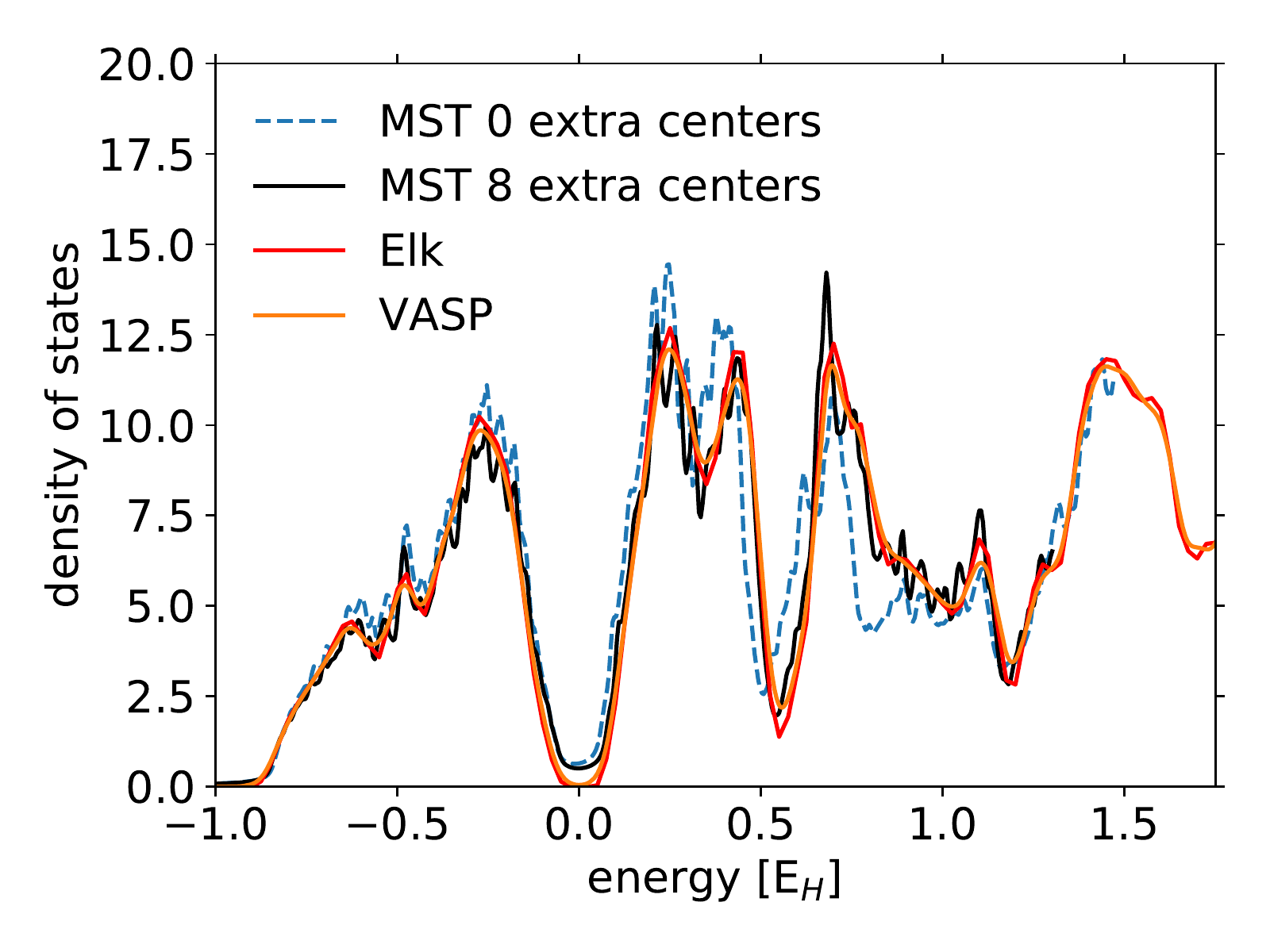}
\caption{Density of states (DOS) for diamond at a temperature of 0.5 eV.  VASP and Elk calculations (present work) are compared to MST calculations.  The MST calculation with 8 extra expansion centers noticeably improves agreement with the VASP and Elk results, compared to the calculation with 0 extra centers.}
\label{fig:diamond_dos}
\end{figure}
\begin{figure}
\centering
\includegraphics[scale=.37]{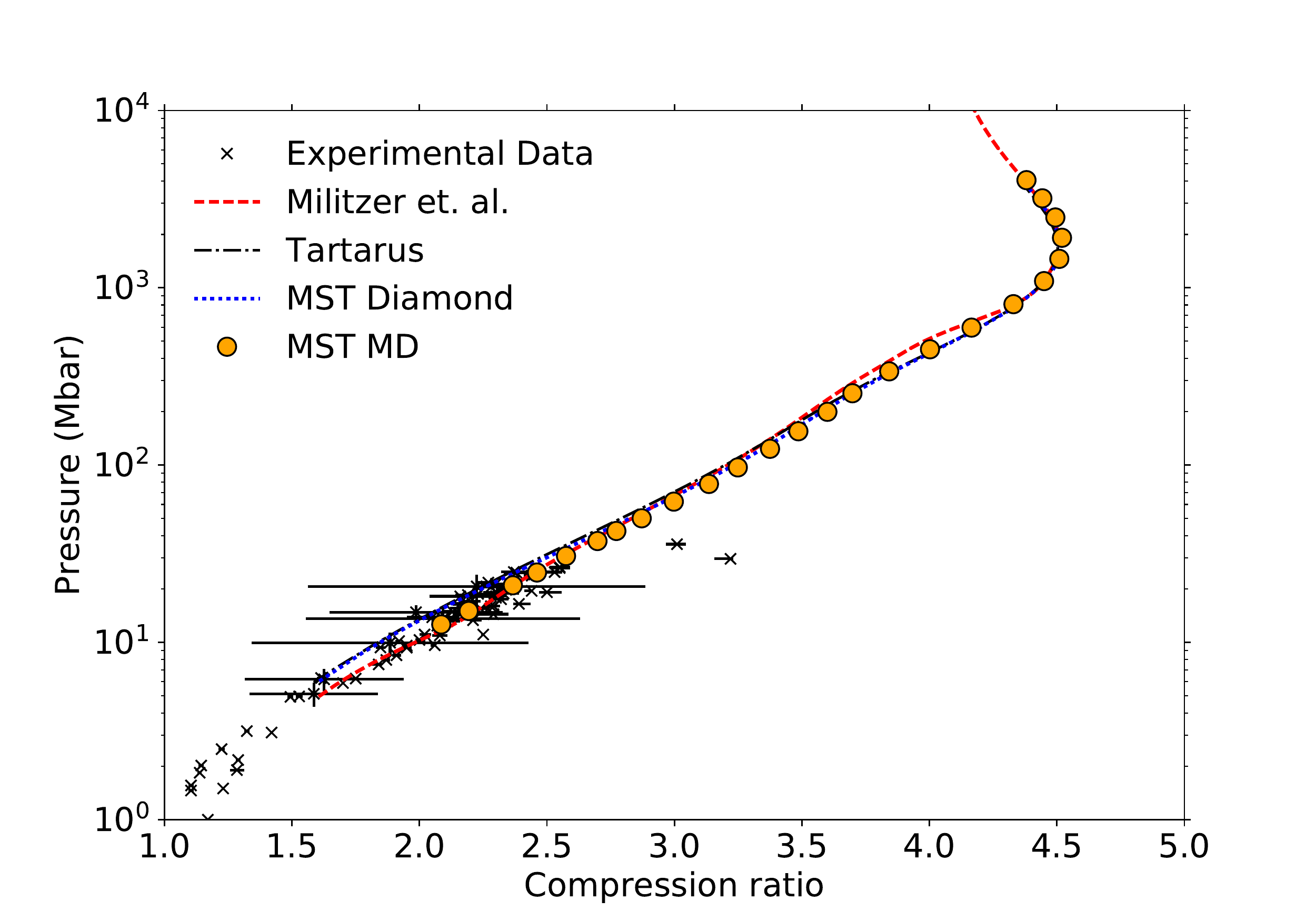}
\caption{Diamond Hugoniot with experimental data \cite{nagao_2006,pavlovskii_1971,kondo_1971, gregor_2017, millot_2020, katagiri_2020, mcwilliams_2010, rus_eos_database}.  Shown are results from the present MST calculation with an assumed diamond structure (MST Diamond), and the present model with molecular dynamics configurations (MST MD), as well as the FPEOS model \cite{militzer_fpeos}, and the \texttt{Tartarus} average atom model \cite{starrett2019wide}.  Note that the initial density was taken to be 3.52 g/cm$^3$.}
\label{fig:c_hug_w_exp_data}
\end{figure}
\begin{figure}
\centering
\includegraphics[scale=.55]{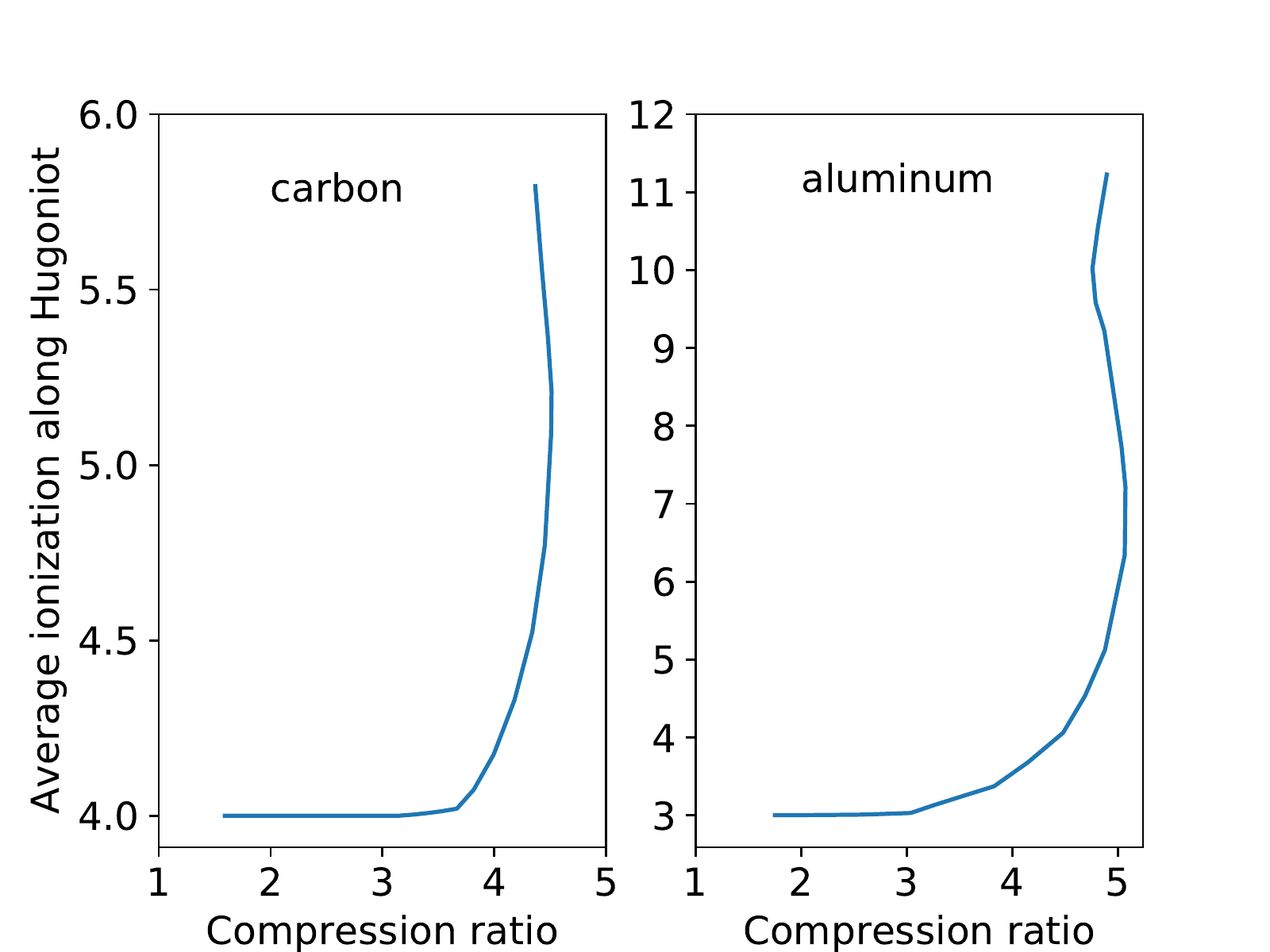}
\caption{Average ionization along principal Hugoniot from the \texttt{Tartarus} average atom model.}
\label{fig:zbar}
\end{figure}
Next, we consider a simplified model, labeled ``No MS" in figure \ref{fig:al_hug_w_exp_data}.  This model is identical to the full MST MD calculation except that we set the multiple scattering Green's function, equation (\ref{gfms}), to be identically zero.
Thus, due to its relative computational simplicity and rapidity, it is a useful 1$^{st}$ approximation for including ionic structure, and as such, represents an intermediate model between average atom and the full multiple scattering solution.  Indeed, it agrees rather well with the full multiple scattering Hugoniot, figure \ref{fig:al_hug_w_exp_data}, showing the stiffer feature for compressions of 2.5 to 4.5 that is seen in the full calculation.  It does not agree perfectly however, and in particular, disagrees more for lower pressures (temperatures), where multiple scattering is more important.

Finally, we show comparison with the FPEOS calculation of Militzer et al., \cite{militzer_al,militzer_fpeos}.  The FPEOS approach uses plane-wave based density functional theory (DFT) molecular dynamics based calculations up to ~174 eV ($\sim$700 Mbar), and then switches to Path-Integral Monte Carlo (PIMC) calculations for higher temperatures.  There is good agreement between this method and MST MD for pressures below 70 Mbar, and above 1 Gbar.  In the region in between, corresponding to the ionization of the $n=2$ shell, the FPEOS Hugoniot is significantly stiffer.  This lobe on the Hugoniot is in the region covered by the DFT calculations in FPEOS.  Both our calculation and FPEOS use 8-atom calculations in this region, and both use an LDA approximation for the exchange and correlation.  We have used the temperature-dependent LDA of Karasiev et al \cite{karasiev14}, whereas the FPEOS use the zero temperature LDA of \cite{ceperley80ground}.   We have recalculated the {\texttt Tartarus} Hugoniot using a zero-temperature LDA \cite{perdew1981self}, and found no significant difference (supplemental material).  Since the calculations both use DFT simulations of the same size, and the difference is not due to the exchange and correlation functional, this leaves four possible sources for this discrepancy: 1) Difference due to nuclear positions, 2) some ill-converged numerical parameter, 3) the muffin-tin approximation in MST, 4) the pseudo-potential approximation in FPEOS.

Let us address these in turn.  1) The good agreement of MST with the fcc calculation for the $n=2$ feature strongly indicates that this should not be the source of the difference, provided reasonable MD configurations are used.  2) We have checked that our calculation is converged in terms of the number of energy grid points, the correlation radius, the number of extra centers, and other checks.  3) The muffin-tin approximation should perform more poorly at lower temperatures, where electrons have less energy on average and are therefore more sensitive to the details of the potential.  As the differences shows up that these rather high temperatures ($>$ 40 eV), it is unlikely that the muffin-tin approximation is to blame.  4) The pseudo-potential approximation that appears in the plane-wave calculations does affect core states and has been the source of numerical issues in other calculations. The effect of the pseudo-potential approximation on the EOS of aluminum has been quantified in reference \cite{militzer_al}, however, we are unable to determine if the observed differences are sufficient to cause the differences in the Hugoniot (figure \ref{fig:al_hug_w_exp_data}).  

The table in the supplementary data section of reference \cite{militzer_al} gives the EOS from which their Hugoniot curve was calculated.   At a temperature of $\sim$174.2 eV and compression of 4.5, they report energies and pressure from both their calculation methods: PIMC and DFT-MD.  At that temperature they report that DFT-MD has a internal energy $\sim$5.3 E$_H$ lower than PIMC, and a pressure $\sim$1\% lower.  We have checked the effects on the Hugoniot curve such changes would have using our EOS.  For pressure, a reduction of 1\% moves the compression at 50.1 eV from 4.86 to 4.89, and at 158.4 eV from 4.94 to 4.97, i.e. relatively small effects.  
Figure \ref{fig:al_hack} shows the effect of shifting the internal energy on the Hugoniot.  At a temperature of 50.1 eV, a shift of just 1 E$_H$ is sufficient to explain the difference between MST and FPEOS.  At a temperature of 158.4 eV, the two calculations can be reconciled with a shift of 5 E$_H$, which is comparable to the reported differeence between PIMC and DFT-MD at 174.2 eV.

Further, also shown in figure \ref{fig:al_hack}, are our own calculations of Hugoniot points at 50.1 eV using the Elk code version 6.8.4~\cite{elk} and VASP version 5.4.4~\cite{kresse1993ab,kresse1996efficient,kresse1996efficiency}. Both calculations use a fcc crystal structure, adding on the ideal ion pressure and kinetic energy to determine the Hugoniot point. In both codes, we use the LDA of Perdew and Zunger~\cite{perdew1981self}. For Elk calculations, we constructed a species file that treats the 1s states as core states and the 2s and higher energy states as valence states. This species file is constructed with a 0.6 Bohr muffin tin radius in order to ensure that nearest-neighbor atoms in the fcc configuration at the compressions studied do not overlap (nearest neighbor distance is roughly 3.2 Bohr at the levels of compression studied). In Elk, we use the built-in ``very high quality"  parameter set, with the exception that we further increase the default plane-wave cutoff energy to a value corresponding to 5,442 eV. We also use a $12\times12\times12$ Gamma-centered k point grid (72 total irreducible k points) and 300 non-spin polarized unoccupied states to ensure all states with fractional occupations of $10^{-10}$ and above are included in the calculation. Calculations are converged so that the total energy changes by less than $10^{-5}$ Hartree and the Kohn-Sham potential changes by less than $10^{-6}$ Hartree. For VASP calculations, we use the Projector Augmented Wave (PAW) method~\cite{blochl1994projector} to treat the 1s core states, with the 2s and higher energy states treated as valence states. The PAW potential we use was constructed specifically for LDA calculations, referred to as the \texttt{Al\_sv\_GW} PAW LDA potential in VASP~\cite{kresse1999ultrasoft}. This potential is constructed using a 1.7 Bohr PAW radius, which at the compression studied in the fcc configuration allows the PAW radii of nearest neighbors to overlap by up to 7\%. We also use a 3000 eV plane-wave energy cutoff, a $12\times12\times12$ Gamma-centered k-point mesh  (72 total irreducible k points), and include 200 non-spin polarized bands to ensure inclusion of all states with fractional occupations of $10^{-6}$ and above. Calculations are converged so that the total energy changes to less than $10^{-6}$ eV. The Hugoniot points from both VASP and Elk agree well with our MST fcc calculation.  In addition, we find that for fcc aluminum, the energies calculated using VASP, Elk, and MST along the 13.136 g/cm$^3$ isochore, from 1 eV up to 50.1 eV, are in excellent agreement (see Supplemental Material).  Both the agreement in the Hugoniot points and the isochore provide strong verification our code and method.

We have also performed MST calculations for carbon, assuming a diamond initial structure.  The EOS of carbon is of interest due to its relevance to white dwarf modeling \cite{kritcher2020measurement}.  Figure \ref{fig:c_hug_w_exp_data} shows the principal Hugoniot for diamond.  
The crystal structure of diamond in the initial state presents a difficulty because, unlike an fcc structure, it is not close-packed. The expansion of the Green's function that only includes the nuclear centers is therefore inaccurate, in contrast to fcc structures.  As a metric to understand this, the percentage of the volume in the muffin-tin spheres for fcc is 74\% (using only nuclear positions as expansion centers).  For diamond the number is 34\%.  Hence, we add 8 extra expansion centers, filling 68\% of the volume.   This improves agreement of the MST result with our density of states (DOS) calculations for diamond using Elk and VASP (figure \ref{fig:diamond_dos}).  The remaining small differences are due in part to different broadening (0.5 eV for Elk and VASP versus 0.27 eV for MST), and other part due to the muffin-tin approximation. 

The MST MD diamond Hugoniot agrees very well with the FPEOS calculation \cite{driver12, benedict14}.  As for aluminum, the ionization feature due to the 1$s$ shell is in good agreement between the approaches.  The effect due to ionic disorder appears to be smaller than for aluminum, with a slight stiffening of the Hugoniot near a compression of 4.  Differences between the full MST MD result and that assuming a diamond structure are largely confined to lower compressions.  Note that the PAMD method for producing the nuclear positions is reliable for carbon only at elevated temperatures, roughly above 5 eV \cite{starrett14b}, due to a neglect of chemical bonds in that method.  The first four points on our MST MD curve  correspond to temperatures from 1.2 to 3.9 eV, while the 5$^{th}$ point is for a temperature of 6.3 eV.  Hence, the nuclear positions are probably somewhat unrealistic for the first four points, where there is a small disagreement with the FPEOS results. 

To explore further why ionic structure seems to have more influence on the aluminum Hugoniot compared to that for carbon, in figure \ref{fig:zbar} we show the average ionization along these Hugoniots as predicted by the \texttt{Tartarus} average atom model, $\bar{Z}$.  As has been discussed many times, this quantity is not uniquely definable \cite{sterne07,starrett2019wide,murillo13partial}.  Here, we choose to define ionized electrons as those having enough energy to escape the average atom potential.  This includes, for example, electrons in resonance states, but not in bound states.  This definition therefore is sensitive to changes in the states close to this threshold, which are also likely to be sensitive to the ionic environment.  Figure \ref{fig:zbar} shows that for carbon, the $1s^2$ state does not start to ionize until the compression reaches $\sim 3.6$.  This compression corresponds to the slight stiffening of the Hugoniot near a compression of 4 mentioned above (figure \ref{fig:c_hug_w_exp_data}).    For aluminum $\bar{Z} \sim 3$ up to a compression of $\sim 3$, then ionization of the $n=2$ shell begins.  Again, this corresponds to the stiffening of the MST MD Hugoniot in figure \ref{fig:al_hug_w_exp_data}.  We therefore conclude that the Hugoniot is sensitive to ionic structure where shell ionization is beginning, for the two cases we have studied, and this effect is expected for other materials.  The size of the effect is larger for aluminum, presumably because the $n=2$ shell contains 8 electrons before it is ionized (out of a total of 13 per atom), whereas for carbon, the $n=1$ shell contains only 2 out of 6.  The observed effects on the Hugoniots due to a disordered ionic environment are confined to this weakly ionized-shell region, and do not extend to higher pressures.  This is due to the increased level of ionization being less sensitive to the ionic environment as these electrons are promoted into higher lying energy states.

\section{Conclusions}
We have presented calculations of principal Hugoniots for diamond and aluminum using Multiple Scattering Theory (MST).  Results including ionic disorder, through the use of molecular dynamics, as well as for fixed crystal structures, were given.  It was found that ionic disorder most strongly affects the Hugoniot curve where bound states are beginning to be ionized.  But for higher pressures (temperatures) still, the effect of ionic disorder is overwhelmed by the higher level of ionization and increased thermal energy of the particles, which washes out the more subtle influence of the ionic disorder on the states near the ionization threshold.

Comparison was also made the the FPEOS model \cite{militzer_fpeos}, and generally good agreement was found.  The one instance of significant disagreement, was the $n=2$ shell ionization feature for aluminum.  We have argued that this could be caused by the precision of the FPEOS reported results for that case.

\section*{Acknowledgments}
The authors thank Dr. N. Shaffer for helpful conversations.
This work was performed under the auspices of the United States Department of Energy under contract DE-AC52-06NA25396.

\bibliographystyle{unsrt}
\bibliography{phys_bib}
\end{document}


\title{Effect of Ionic Disorder on the Principal Shock Hugoniot - Supplementary Material}
\author{Crystal Ottoway}
\author{Daniel A. Rehn}
\author{Didier Saumon}
\author{C. E. Starrett}
\email{starrett@lanl.gov}

\affiliation{Los Alamos National Laboratory, P.O. Box 1663, Los Alamos, NM 87545, U.S.A.}
\maketitle

\section{Addition comparisons and data}
\begin{figure}
\centering
\includegraphics[scale=.55]{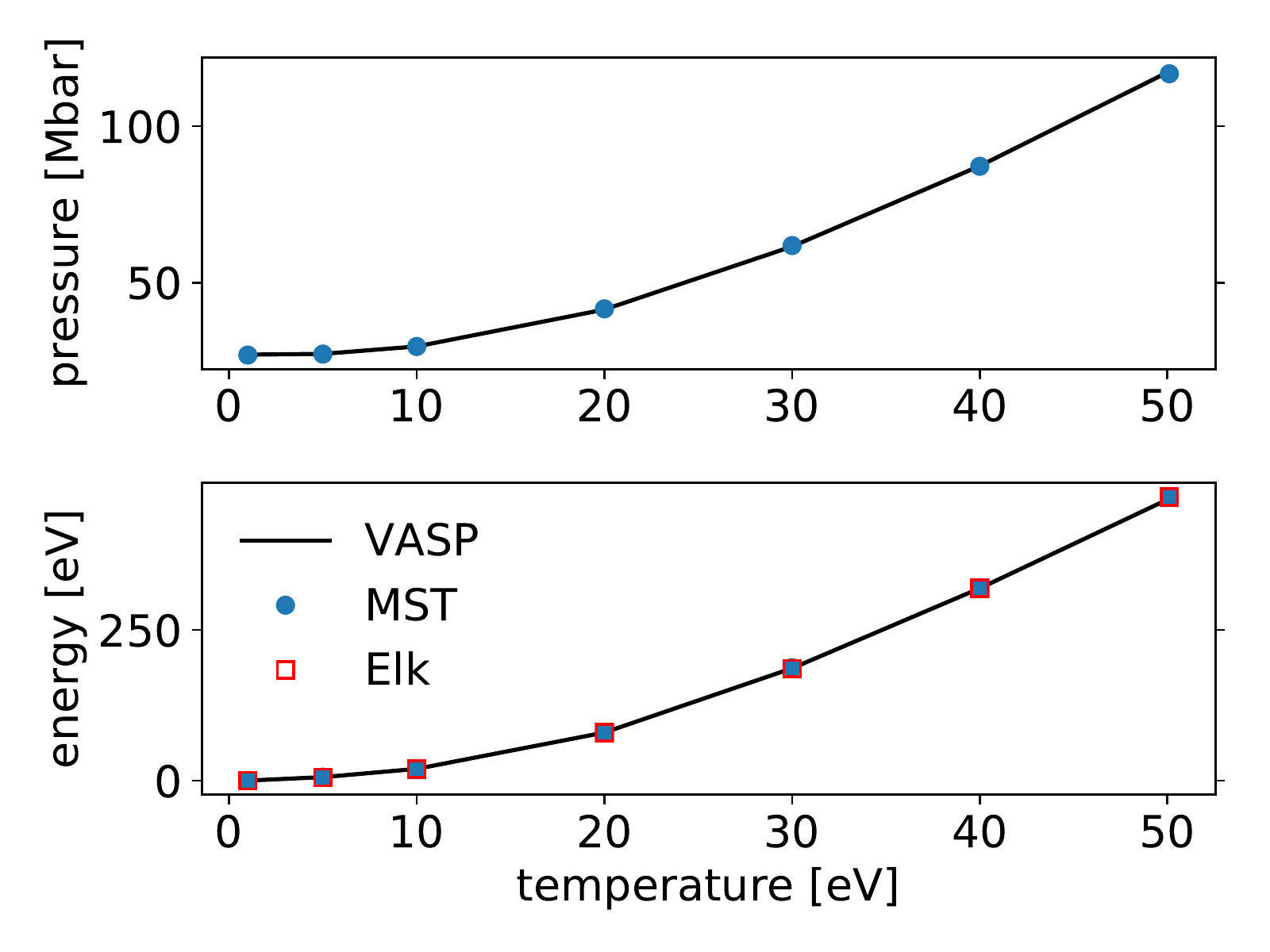}
\caption{Equation of state of fcc aluminum along a 13.136 g/cm$^3$ isochore.  We compare results from the present MST model to the Elk \cite{elk} and VASP \cite{kresse1993ab,kresse1996efficiency,kresse1996efficient} codes.}
\label{fig:eosfcc}
\end{figure}

In figure \ref{fig:eosfcc} we compare our Multiple Scattering Theory (MST) results for the EOS of fcc aluminum at 13.136 g/cm$^3$ to those from the Elk and VASP codes.  Note that these do not include terms for the kinetic energy of the nuclei.  Agreement is excellent across all three codes for energy.  For pressure, the agreement between VASP and MST is again very good.  Note that the Elk code does not return the pressure, so we have not reported it.  For the Hugoniot point in the main paper, we have calculated pressure from Elk using a finite difference approximation to $P=-\frac{dF}{dV}$, where $P$ is the pressure, $F$ is the free energy and $V$ is the volume.  

In tables \ref{tab:alhug} and \ref{tab:chug} we give the Hugoniot points from the MST calculations for aluminum and carbon. 

\begin{figure}
\centering
\includegraphics[scale=.55]{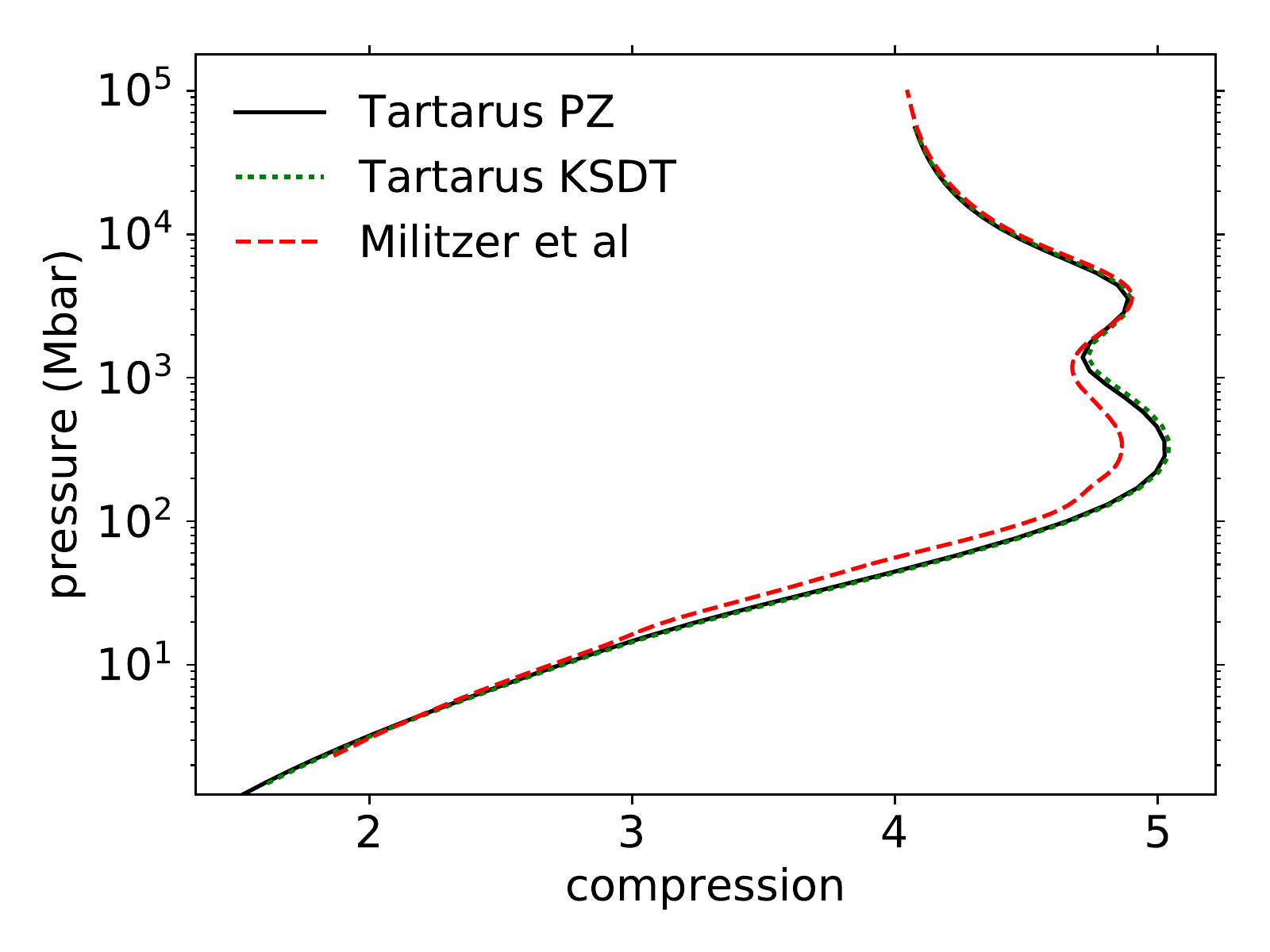}
\caption{The effect of a temperature dependent LDA, KSDT \cite{ksdt}, versus a temperature-independent LDA \cite{perdew1981self}, on the Hugoniot of aluminum.  Little difference is observed and it does not explain the difference with the result of Militzer et al \cite{militzer_al}.}
\label{fig:tartarus}
\end{figure}

In figure \ref{fig:tartarus} we show the effect of using a temperature-dependent LDA approximation \cite{ksdt}, versus a zero-temperature LDA \cite{perdew1981self}.  The small difference observed is insufficient to explain the difference with the FPEOS calculation of Militzer et al \cite{militzer_fpeos}.


\begin{table}[]
    \centering
    \begin{tabular}{ccc}
        \hline
Temperature (eV) & Compression &Pressure (Mbar) \\
\hline
\\
1 &1.9545 &2.6582 \\
1.2 &2.0014 &2.9159\\
1.9 &2.1650 &3.9287\\
2.5 &2.2853 &4.8470\\
3.1 &2.3783 &5.6130\\
3.9 &2.4871 &6.6965\\
5 &2.6089 &8.5324\\
6.3 &2.7594 &10.3668\\
7.9 &2.8538 &12.7977\\
10 &2.9957 &16.6297\\
12.5 &3.1408 &20.9275\\
15.8 &3.3206 &28.3041\\
19.9 &3.5738 &38.1270\\
25.1 &3.9122 &52.7835\\
31.6 &4.2971 &72.2960\\
39.8 &4.6622 &99.3885\\
50.1 &4.8487 &139.4969\\
63.1 &4.9692 &194.5437\\
79.4 &5.0372 &267.5213\\
100 &5.0491 &364.8706\\
125.8 &5.0210 &492.2226\\
158.4 &4.9534 &656.8414\\
199.5 &4.8672 &867.0317\\
251.1 &4.7866 &1135.6739\\
316.2 &4.7566 &1495.3849\\
398.1 &4.8082 &2003.4752\\
500 &4.8908 &2703.7956\\
    \end{tabular}
    \caption{MST Hugoniot points for aluminum.}
    \label{tab:alhug}
\end{table}

\begin{table}[]
    \centering
    \begin{tabular}{ccc}
        \hline
Temperature (eV) & Compression &Pressure (Mbar) \\
\hline
\\
1.2	&2.0854	&12.6286\\
1.9	&2.1929	&15.0374\\
3.1	&2.3662	&20.9965\\
3.9	&2.4608	&24.7841\\
5	&2.5748	&30.7061\\
6.3	&2.6981	&37.3048\\
7.9	&2.7721	&42.5023\\
10	&2.8716	&50.1097\\
12.5 &2.9975 &62.2310\\
15.8 &3.1353 &78.2810\\
19.9 &3.2486 &97.0429\\
25.1 &3.3749 &123.5633\\
31.6 &3.4860 &155.1837\\
39.8 &3.5999 &199.7690\\
50.1 &3.6980 &254.2529\\
63.1 &3.8422 &337.8647\\
79.4 &4.0023 &450.0254\\
100	&4.1652	&596.3107\\
125.8 &4.3295 &808.5757\\
158.4 &4.4496 &1091.0550\\
199.5 &4.5094 &1456.4925\\
251.1 &4.5199 &1915.2162\\
316.2 &4.4939 &2493.8412\\
398.1 &4.4429 &3197.1812\\
500.0 &4.38057 &4048.8529\\
    \end{tabular}
    \caption{MST Hugoniot points for carbon.}
    \label{tab:chug}
\end{table}

\bibliographystyle{unsrt}
\bibliography{phys_bib}